\PassOptionsToPackage{unicode}{hyperref}
\PassOptionsToPackage{hyphens}{url}
\PassOptionsToPackage{dvipsnames,svgnames,x11names}{xcolor}
\documentclass[
  10pt,
]{article}
\usepackage{amsmath,amssymb}
\usepackage{iftex}
\ifPDFTeX
  \usepackage[T1]{fontenc}
  \usepackage[utf8]{inputenc}
  \usepackage{textcomp} 
\else 
  \usepackage{unicode-math} 
  \defaultfontfeatures{Scale=MatchLowercase}
  \defaultfontfeatures[\rmfamily]{Ligatures=TeX,Scale=1}
\fi
\usepackage{lmodern}
\ifPDFTeX\else
\fi
\IfFileExists{upquote.sty}{\usepackage{upquote}}{}
\IfFileExists{microtype.sty}{
  \usepackage[]{microtype}
  \UseMicrotypeSet[protrusion]{basicmath} 
}{}
\makeatletter
\@ifundefined{KOMAClassName}{
  \IfFileExists{parskip.sty}{%
    \usepackage{parskip}
  }{
    \setlength{\parindent}{0pt}
    \setlength{\parskip}{6pt plus 2pt minus 1pt}}
}{
  \KOMAoptions{parskip=half}}
\makeatother
\usepackage{xcolor}
\usepackage{graphicx}
\makeatletter
\newsavebox\pandoc@box
\newcommand*\pandocbounded[1]{
  \sbox\pandoc@box{#1}%
  \Gscale@div\@tempa{\textheight}{\dimexpr\ht\pandoc@box+\dp\pandoc@box\relax}%
  \Gscale@div\@tempb{\linewidth}{\wd\pandoc@box}%
  \ifdim\@tempb\p@<\@tempa\p@\let\@tempa\@tempb\fi
  \ifdim\@tempa\p@<\p@\scalebox{\@tempa}{\usebox\pandoc@box}%
  \else\usebox{\pandoc@box}%
  \fi%
}
\def\fps@figure{htbp}
\makeatother
\NewDocumentCommand\citeproctext{}{}

\makeatletter
 \let\@cite@ofmt\@firstofone
 \def\@biblabel#1{}
 \def\@cite#1#2{{#1\if@tempswa , #2\fi}}
\makeatother
\newlength{\cslhangindent}
\newlength{\csllabelwidth}
\newenvironment{CSLReferences}[2] 
 {\begin{list}{}{%
  \setlength{\itemindent}{0pt}
  \setlength{\leftmargin}{0pt}
  \setlength{\parsep}{0pt}
  \ifodd #1
   \setlength{\leftmargin}{\cslhangindent}
   \setlength{\itemindent}{-1\cslhangindent}
  \fi
  \setlength{\itemsep}{#2\baselineskip}}}
 {\end{list}}
\usepackage{calc}

\newcommand{\CSLLeftMargin}[1]{\parbox[t]{\csllabelwidth}{\strut#1\strut}}
\newcommand{\CSLRightInline}[1]{\parbox[t]{\linewidth - \csllabelwidth}{\strut#1\strut}}

\usepackage{dcolumn, rotating, longtable, lineno, float, array, tabularx, inputenc}

\graphicspath{{figures/}}
\usepackage[margin=2.5cm]{geometry}
\usepackage{setspace}
\usepackage{booktabs}
\usepackage{longtable}
\usepackage{array}
\usepackage{multirow}
\usepackage{wrapfig}
\usepackage{float}
\usepackage{colortbl}
\usepackage{pdflscape}
\usepackage{tabu}
\usepackage{threeparttable}
\usepackage{threeparttablex}
\usepackage[normalem]{ulem}
\usepackage{makecell}
\usepackage{xcolor}
\usepackage{bookmark}
\IfFileExists{xurl.sty}{\usepackage{xurl}}{} 
\hypersetup{
  pdftitle={Inequality traps detected in sustainable development goals data},
  colorlinks=true,
  linkcolor={blue},
  filecolor={Maroon},
  citecolor={blue},
  urlcolor={blue},
  pdfcreator={LaTeX via pandoc}}

\title{Inequality traps detected in sustainable development goals data}
\author{\small Juan C. Rocha\textsuperscript{1,2}, Maike
Hamann\textsuperscript{3}, Jiangxiao Qiu\textsuperscript{4}, Tong
Wu\textsuperscript{5}, Tomas Chaigneau\textsuperscript{3}, Emilie
Lindkvist\textsuperscript{1},\\
\small  Caroline Schill\textsuperscript{1,2,6}, Alon
Shepon\textsuperscript{7,8}, Andrew R. Tilman\textsuperscript{9},
Geraldine D. Verkleij\textsuperscript{6}, Anne-Sophie
Crépin\textsuperscript{1,6}, and Carl Folke\textsuperscript{1,6}\\
\footnotesize \textsuperscript{1}Stockholm Resilience Centre, Stockholm
University\\
\footnotesize \textsuperscript{2}Anthropocene Laboratory, Royal Swedish
Academy of Sciences\\
\footnotesize \textsuperscript{3}Environment and Sustainability
Institute, University of Exeter\\
\footnotesize \textsuperscript{4}School of Forest, Fisheries, and
Geomatics Sciences, Fort Lauderdale Research and Education Center,\\
\footnotesize Institute of Food and Agricultural Sciences, University of
Florida\\
\footnotesize \textsuperscript{5}The Natural Capital Project and
Institute for International Studies, Stanford University\\
\footnotesize \textsuperscript{6}Beijer Institute of Ecological
Economics, Royal Swedish Academy of Sciences\\
\footnotesize \textsuperscript{7}The Steinhardt Museum of Natural
History, Tel Aviv University, Tel Aviv\\
\footnotesize \textsuperscript{8}The Department of Public Policy, Social
Sciences Faculty, Tel Aviv University, Tel Aviv\\
\footnotesize \textsuperscript{9}USDA Forest Service, Northern Research
Station, St.~Paul, MN, USA\\
\small \texttt{\href{mailto:juan.rocha@su.se}{\nolinkurl{juan.rocha@su.se}}}}
\date{}

\begin{document}
\maketitle

\begin{abstract}

  The relationship between inequality and the biosphere has been hypothesized to mutual dependecies and feedbacks. If that is true, such feedbacks may give rise to inequality regimes and potential tipping points between them. Here we explore synergies and trade-offs between inequality and biosphere-related sustainable development goals. We used the openly available SDG datasets by the World Bank (WB) and United Nations (UN) and applied ordination methods to distill interactions between economic inequality and the environmental impact across countries. Our results confirm the existence of inequality regimes, and we find preliminary evidence that corruption may be a candidate driver of tipping between regimes.

\end{abstract}

\section{Introduction}\label{introduction}

Countries around the world have committed to achieve 17 sustainable
development goals (SDGs) as proposed by the United Nations. The
ambitious agenda is materialized in 169 targets and indicators, yet not
all targets are monitored or properly measured, and not all countries
report them, presenting significant data and knowledge
gaps\textsuperscript{1}. An open question in sustainability science is
whether these targets are simultaneously achievable or if trade-offs
occur between them? Previous work on synergies and trade-offs between
SDGs suggest that both may be present\textsuperscript{2--5}. Yet this
work is based on expert elicitation and correlational studies, which are
limited in their ability to identify path dependencies in development
trajectories or mechanistic relationships among indicators. As a result,
it remains unclear whether structural constraints limits a country's
capacity to achieve the SDGs. Relatedly, it may be the case that certain
milestones must be unlocked before development can proceed towards the
achievement of a particular dimension of sustainability. These
complexities suggest that new analyses may be necessary to guide
policymakers toward simultaneous achievement of the SDGs

When the SDGs were adopted in 2015, the United Nations created a
comprehensive, annually-updated database of SDG
indicators\textsuperscript{2,3}. Based on these data, previous analyses
found that goals associated with poverty alleviation, well-being,
economic development, and innovation (e.g.~SDGs 1, 3, 7, 8, 9) tended to
synergize with other goals, while goals related to responsible
consumption, climate action, natural resources, and cooperation showed
the most trade-offs (e.g.~SDGs 11, 12, 13, 14, 16,
17)\textsuperscript{4,5}. Drawing upon relevant World Bank data, Lusseau
and Mancini\textsuperscript{6} find these patterns to be modulated by
countries' overall income level, with low-income countries showing
synergies across all goals, while trade-offs start to appear in higher
income countries. More recently, Xiao \emph{et al.},\textsuperscript{7}
analysed transboundary SDG interactions and found that high income
countries play a disproportionate role in influencing the achievement of
SDGs in other countries. These differences suggest that it may be
worthwhile investigating if the nature of the mechanisms linking SDGs to
each other could depend on income inequality or other factors that
modulate it, such as corruption (measured by the corruption perception
index\textsuperscript{8})\textsuperscript{9}.

Here we explore the possible existence of inequality traps and regimes
by investigating synergies and trade-offs between inequality and
biosphere-related sustainable development goals across nations (SDGs 5,
6, 10, 13, 14 and 15). The motivation is threefold. First, the
Convention for Biological Diversity is currently negotiating and
agreeing on the next set of goals and ambitions to mitigate biodiversity
loss. A deeper understanding of the relationships among sustainable
development indicators may bolster progress towards this vision by
enabling the setting of realistic, achievable targets for all nations,
regardless of their current development trajectory\textsuperscript{10}.
Second, within-country inequality has been rising in the last decades
even in high income countries\textsuperscript{11}, and as a consequence
of the Covid pandemic, in-between-country inequality has risen for the
first time in a generation\textsuperscript{12}. These statistics set
back progress on the inequality SDG by at least a
decade\textsuperscript{13}, and undermines the mantra of leaving no one
behind. Lastly, recent conceptual and theoretical work has proposed
mechanisms by which an increase in inequality can impact the
environment, while changes in the environment can feedback and further
impact inequalities\textsuperscript{14--16}.

Previous work on the origins and persistence of inequality have proposed
mechanisms across scales. For example, micro-level dynamics such as
aspirations, conspicuous consumption, social norms, and perceptions of
fairness have been proposed as potential mechanisms linking inequality
and the biosphere by disincentivising cooperation\textsuperscript{14}.
At the meso-level, market-concentration and lobbying have been proposed
as mechanisms by which powerful actors tend to favor institutions that
further enable capital accumulation\textsuperscript{14,17}. At the
national scale, tax policies are key mechanisms for redistribution, but
they are not easily comparable across countries\textsuperscript{18}.
Another key mechanism proposed is corruption\textsuperscript{8,9},
which, similar to market concentration, enables actors in power to seek
their individual interests at the expense of the social good.

Recent work also shows that a trilemma exists where countries struggle
to simultaneously achieve high prosperity, high environmental standards,
while reducing inequality\textsuperscript{19}. Using data from
environmental footprint, the gross national product, and the Gini
coefficient time series, 53 countries were clustered and a typology of
trajectories identified. No country achieved the three goals
simultaneously, and Latin American countries seem to exhibit dynamics of
an inequality trap or a high inequality regime. Some countries'
development trajectories suggest that social progress can be achieved
without compromising the biosphere\textsuperscript{19}. While no country
has simultaneously achieved these three goals, some countries are indeed
moving in the right direction\textsuperscript{19--21}.

It remains an open question whether these patterns are robust across
different datasets and scales, or whether there exist specific driving
factors and feedback mechanisms that underlie inequality traps. The
dichotomy between low-income countries exhibiting synergies across all
goals, while trade-offs start to appear in higher-income
countries\textsuperscript{6} motivates the need to study mechanisms
explaining these trade-offs and how they may differ due to countries'
income level. If the hypothesis of the inequality trap is true, we
should observe bimodal or multimodal distributions across inequality and
environmental variables. Each mode would correspond to an inequality
regime, and the transitions probability of staying within one regime
should be much higher than the probability of shifting regimes. If there
are nonlinear dynamics in inequality keeping countries trapped in a
particular regime, then we could also observe hysteresis or different
break points between regimes. Here we explore the possible existence of
inequality traps and regimes by investigating synergies and trade-offs
between inequality and biosphere-related sustainable development goals
across nations.

\section{Methods}\label{methods}

\emph{Datasets:} We used the SDGs datasets made openly available by the
World Bank (WB) and United Nations (UN). The WB dataset offers 403
indicators, with time series from 1990 to 2019 for 263 countries or
administrative areas (N = 2 013 791 observations). The UN dataset offers
time series from 1963 to 2025 for 17 SDGs, 168 targets, and 247
indicators, 687 time series, in 413 administrative areas (N = 2 821 669
observations). Despite their coverage, both datasets contain a high
proportion of missing values, some countries have better temporal
coverage than others. We focused our analysis on country level
indicators only for SDGs 5, 6, 10, 13, 14 and 15 that relate to
inequality and the biosphere. We complemented the SDGs datasets with
inequality data from the World Inequality database (WID), using their
estimates of the ratio of pre-tax national income for working adults
(population \textgreater{} 20 years old) computed as the share of the
top 10\% divided by the the share of the bottom 50\%
(\texttt{rptinc992j\_p0p100}), the share of the 1\%
(\texttt{sptinc992j\_p99p100}), and the Gini coefficient
(\texttt{gptinc992j\_p0p100}). We also used their estimates for net
wealth inequality computed as the share of the top 10\% over the share
of the bottom 50\% (\texttt{rhweal992j\_p0p100}), the share of the 1\%
(\texttt{shweal992j\_p99p100}), and the Gini
coefficient(\texttt{ghweal992j\_p0p100}). We also used data from the
quality of government dataset\textsuperscript{8} to investigate the
relationship between inequality indicators and the corruption perception
index.

\emph{Variable selection:} We computed the proportion of missing values
for all time series related to our initial selection of indicators (Figs
\ref{fig:SM_WB_env}, \ref{fig:SM_WB_env2}, \ref{fig:SM_WB_select},
\ref{fig:SM_UNdata}, \ref{fig:SM_UNselect}, \ref{fig:SM_UNred}). We
discarded indicators for which time series contained more than 30\% of
missing values, or less than 45 countries. Missing values were then
imputed using a cubic spline, leaving us with 160 countries, 19 years of
data across 9 indicators for the WB dataset; and 19 series capturing 9
indicators, 68 countries over 22 years for the UN dataset. The UN
dataset was further reduced to 67 countries because the WID does not
report inequality time series for Fiji. Table \ref{tab:tbl1} summarizes
our selected variables, their units and available ranges. A list of the
countries analysed is presented in the supplementary information (SI).

\emph{Ordination:} We used multiple factor analysis (MFA) and principal
component analysis (PCA) to reduce the dimensionality of the data and
explore similarities and differences across countries. MFA enables us to
specify the nested structure of our data and account for repeated
observations of our variables over time. We recovered some of the
qualitative results with PCA as robustness check, but these results are
presented in the SI. We also performed a clustering sensitivity analysis
following the protocols by Charrad\textsuperscript{22} and
Brock\textsuperscript{23}. We tested over 10 clustering techniques and
compared them across \textgreater30 performance metrics to infer from
the data what are the optimal numbers of clusters to fit and preferable
algorithms. The ordination step helped us identify variables with enough
variability and carrying information on inequality or the environment to
explore the next steps of the analysis. The robustness checks on
clustering were necessary to avoid spurious results (e.g.~higher number
of clusters, over fitting) due to the choice of clustering technique or
idiosyncrasies of the data (e.g.~raw distributions).

\emph{Analysis of trajectories:} With the results from the MFA we
identified candidate variables where synergies or trade-offs are
observed. A trade-off in the reduced dimensional space occurs when
improving on the direction of one indicator (e.g.~reducing inequality)
implies a decline in the direction of another indicator over time.
Similarly, a synergy would be when progress in one indicator coincides
with improvement in another indicator. We studied country trajectories
for some of these candidate variables where we found enough variability
to test for bimodality. Density plots helped us identify candidate
variables and expose the main regimes. In the next step, we evaluated
the modality of the distributions of the inequality and biosphere-based
measures. If the hypothesis of inequality traps or inequality regimes is
true, we would expect to find multimodal distributions. We applied the
Hartigan's Dip Test for unimodality. If the test is positive at 1\%
significance level we rejected the hypothesis of unimodality. We also
expect the modes of these distributions to be correlated to the country
groups identified via clustering analysis to discard the possibility
that several modes exist independent (or only partially overlapping)
with the identified country clusters. If the groups are well mixed
between modes, then the multimodal pattern could be the consequence of
some other process (e.g.~seasonality) and not regime shift dynamics. We
further explore the association between inequality regimes and
corruption as a driver through linear regression models. Last, we expect
most countries' trajectories to remain within a single regime in the
parameter space and perhaps a few of them to move between regimes.

\begin{table}[!h]
\centering
\caption{\label{tab:tbl1}Summary of variables used}
\centering
\fontsize{8}{10}\selectfont
\begin{tabu} to \linewidth {>{\raggedright\arraybackslash}p{5mm}>{\raggedleft\arraybackslash}p{5mm}>{\raggedright\arraybackslash}p{25mm}>{\raggedright\arraybackslash}p{8cm}>{\raggedright}X}
\toprule
Source & Goal & Series & Variable & Units\\
\midrule
\cellcolor{gray!10}{UN} & \cellcolor{gray!10}{5} & \cellcolor{gray!10}{SG\_GEN\_PARLN} & \cellcolor{gray!10}{Number of seats held by women in national parliaments} & \cellcolor{gray!10}{number}\\
UN & 5 & SG\_GEN\_PARL & Proportion of seats held by women in national parliaments & \% of total number of seats\\
\cellcolor{gray!10}{UN} & \cellcolor{gray!10}{6} & \cellcolor{gray!10}{SH\_SAN\_SAFE} & \cellcolor{gray!10}{Proportion of population using safely managed sanitation services, by urban/rural} & \cellcolor{gray!10}{0 to 1}\\
UN & 6 & SH\_SAN\_DEFECT & Proportion of population practicing open defecation, by urban/rural & 0 to 1\\
\cellcolor{gray!10}{UN} & \cellcolor{gray!10}{6} & \cellcolor{gray!10}{ER\_H2O\_WUEYST} & \cellcolor{gray!10}{Water Use Efficiency} & \cellcolor{gray!10}{US\$ per cubic meter}\\
\addlinespace
UN & 6 & ER\_H2O\_STRESS & Level of water stress: freshwater withdrawal as a proportion of available freshwater resources & 0 to 1\\
\cellcolor{gray!10}{UN} & \cellcolor{gray!10}{6} & \cellcolor{gray!10}{EN\_LKRV\_PWAN} & \cellcolor{gray!10}{Lakes and rivers permanent water area} & \cellcolor{gray!10}{sq. km}\\
UN & 6 & EN\_LKRV\_PWAP & Lakes and rivers permanent water area & \% of total land area\\
\cellcolor{gray!10}{UN} & \cellcolor{gray!10}{6} & \cellcolor{gray!10}{EN\_LKRV\_SWAN} & \cellcolor{gray!10}{Lakes and rivers seasonal water area} & \cellcolor{gray!10}{sq. km}\\
UN & 6 & EN\_LKRV\_SWAP & Lakes and rivers seasonal water area & \% of total land area\\
\addlinespace
\cellcolor{gray!10}{UN} & \cellcolor{gray!10}{6} & \cellcolor{gray!10}{EN\_LKRV\_PWAC} & \cellcolor{gray!10}{Lakes and rivers permanent water area change} & \cellcolor{gray!10}{NA}\\
UN & 6 & EN\_LKRV\_SWAC & Lakes and rivers seasonal water area change & NA\\
\cellcolor{gray!10}{UN} & \cellcolor{gray!10}{6} & \cellcolor{gray!10}{EN\_RSRV\_MNWAN} & \cellcolor{gray!10}{Reservoir minimum water area} & \cellcolor{gray!10}{sq. km}\\
UN & 6 & EN\_RSRV\_MNWAP & Reservoir minimum water area & \% of total land area\\
\cellcolor{gray!10}{WB} & \cellcolor{gray!10}{10} & \cellcolor{gray!10}{SP.URB.TOTL.IN.ZS} & \cellcolor{gray!10}{Urban population} & \cellcolor{gray!10}{\% of total population}\\
\addlinespace
WB & 10 & EG.ELC.ACCS.ZS & Access to electricity & \% of population\\
\cellcolor{gray!10}{WB} & \cellcolor{gray!10}{10} & \cellcolor{gray!10}{SG.LAW.INDX} & \cellcolor{gray!10}{Women Business and the Law Index Score} & \cellcolor{gray!10}{scale 1-100}\\
WB & 10 & IT.NET.USER.ZS & Individuals using the Internet & 0 to 1\\
\cellcolor{gray!10}{UN} & \cellcolor{gray!10}{10} & \cellcolor{gray!10}{SM\_POP\_REFG\_OR} & \cellcolor{gray!10}{Number of refugees per 100,000 population, by country of origin} & \cellcolor{gray!10}{0 to 1}\\
WII & 10 & rptinc992j\_p0p100 & Ratio of pre-tax national income for working adults & 0 to 1\\
\addlinespace
\cellcolor{gray!10}{WII} & \cellcolor{gray!10}{10} & \cellcolor{gray!10}{rhweal992j\_p0p100} & \cellcolor{gray!10}{Ratio of wealth for working adults} & \cellcolor{gray!10}{0 to 1}\\
WII & 10 & shweal992j\_p99p100 & Share of the top 1\% of wealth & 0 to 1\\
\cellcolor{gray!10}{WII} & \cellcolor{gray!10}{10} & \cellcolor{gray!10}{ghweal992j\_p0p100} & \cellcolor{gray!10}{Gini coefficient of wealth} & \cellcolor{gray!10}{0 to 1}\\
WII & 10 & gptinc992j\_p0p100 & Gini coefficeint of pre-tax income for working adults & NA\\
\cellcolor{gray!10}{WII} & \cellcolor{gray!10}{10} & \cellcolor{gray!10}{sptinc992j\_p99p100} & \cellcolor{gray!10}{Share of the 1\% of pre-tax income for working adults} & \cellcolor{gray!10}{NA}\\
\addlinespace
WB & 13 & EN.ATM.CO2E.PC & CO2 emissions & metric tons per capita\\
\cellcolor{gray!10}{WB} & \cellcolor{gray!10}{13} & \cellcolor{gray!10}{EG.EGY.PRIM.PP.KD} & \cellcolor{gray!10}{Energy intensity level of primary energy} & \cellcolor{gray!10}{MJ/\$2011 PPP GDP}\\
WB & 15 & AG.YLD.CREL.KG & Cereal yield & kg per hectare\\
\cellcolor{gray!10}{WB} & \cellcolor{gray!10}{15} & \cellcolor{gray!10}{AG.LND.FRST.K2} & \cellcolor{gray!10}{Forest area} & \cellcolor{gray!10}{0 to 1}\\
WB & 15 & AG.LND.FRST.ZS & Forest area & 0 to 1\\
\addlinespace
\cellcolor{gray!10}{UN} & \cellcolor{gray!10}{15} & \cellcolor{gray!10}{ER\_PTD\_FRHWTR} & \cellcolor{gray!10}{Average proportion of Freshwater Key Biodiversity Areas (KBAs) covered by protected areas} & \cellcolor{gray!10}{0 to 1}\\
UN & 15 & ER\_PTD\_TERR & Average proportion of Terrestrial Key Biodiversity Areas (KBAs) covered by protected areas & 0 to 1\\
\cellcolor{gray!10}{UN} & \cellcolor{gray!10}{15} & \cellcolor{gray!10}{ER\_PTD\_MTN} & \cellcolor{gray!10}{Average proportion of Mountain Key Biodiversity Areas (KBAs) covered by protected areas} & \cellcolor{gray!10}{NA}\\
UN & 15 & ER\_RSK\_LST & Red List Index & NA\\
\bottomrule
\end{tabu}
\end{table}

\section{Results}\label{results}

We find confirming evidence that inequality regimes exist and that some
countries tend to be trapped on high inequality. Exploring corruption
data as a potential mechanism, we find empirical evidence suggesting the
existence of hysteresis, further providing support for the idea of
potential regime shifts in inequality. We also observe some synergies
and trade-offs between inequality and environmental goals.

\begin{figure*}[ht]
\centering
\includegraphics[width = 7in, height = 4in]{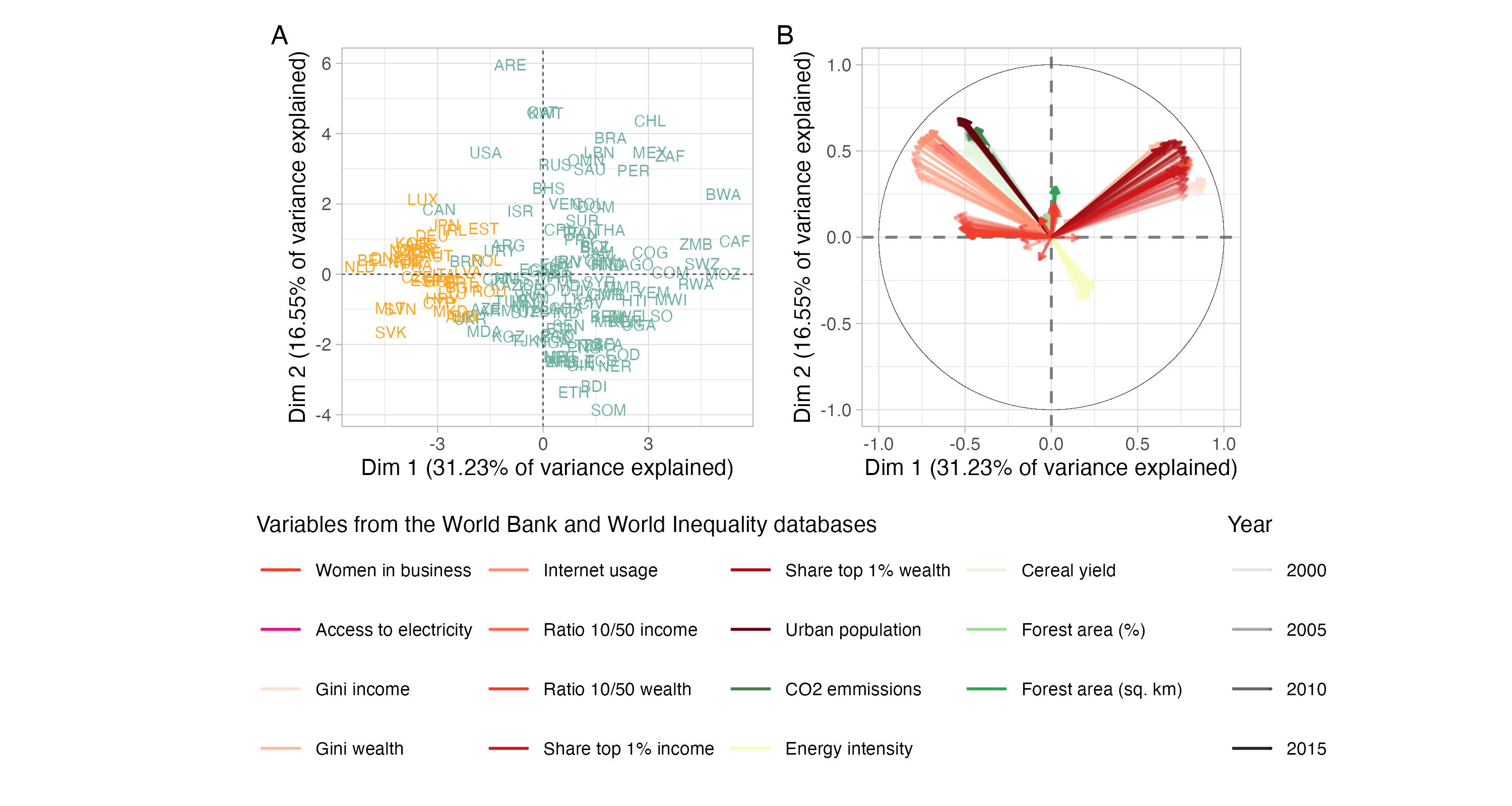}
\caption{\textbf{Multiple factor analysis with World Bank data} The first 10 principal components explain 94.8\% of the variation, the first two (A) explain 53\%. The 10 first components were used to cluster 151 countries resulting in two clusters (A). The correlation circle across explanatory variables is presented in (B) along their loadings on the first two components of the ordination. Variables in the legend are ordered and colored according to the SDGs used (e.g. orange for gender equality, blues for water and sanitation, reds for inequality, and greens for life on land).}
\label{fig:wb}
\end{figure*}

\begin{figure*}[ht]
\centering
\includegraphics[width = 7in, height = 4.5in]{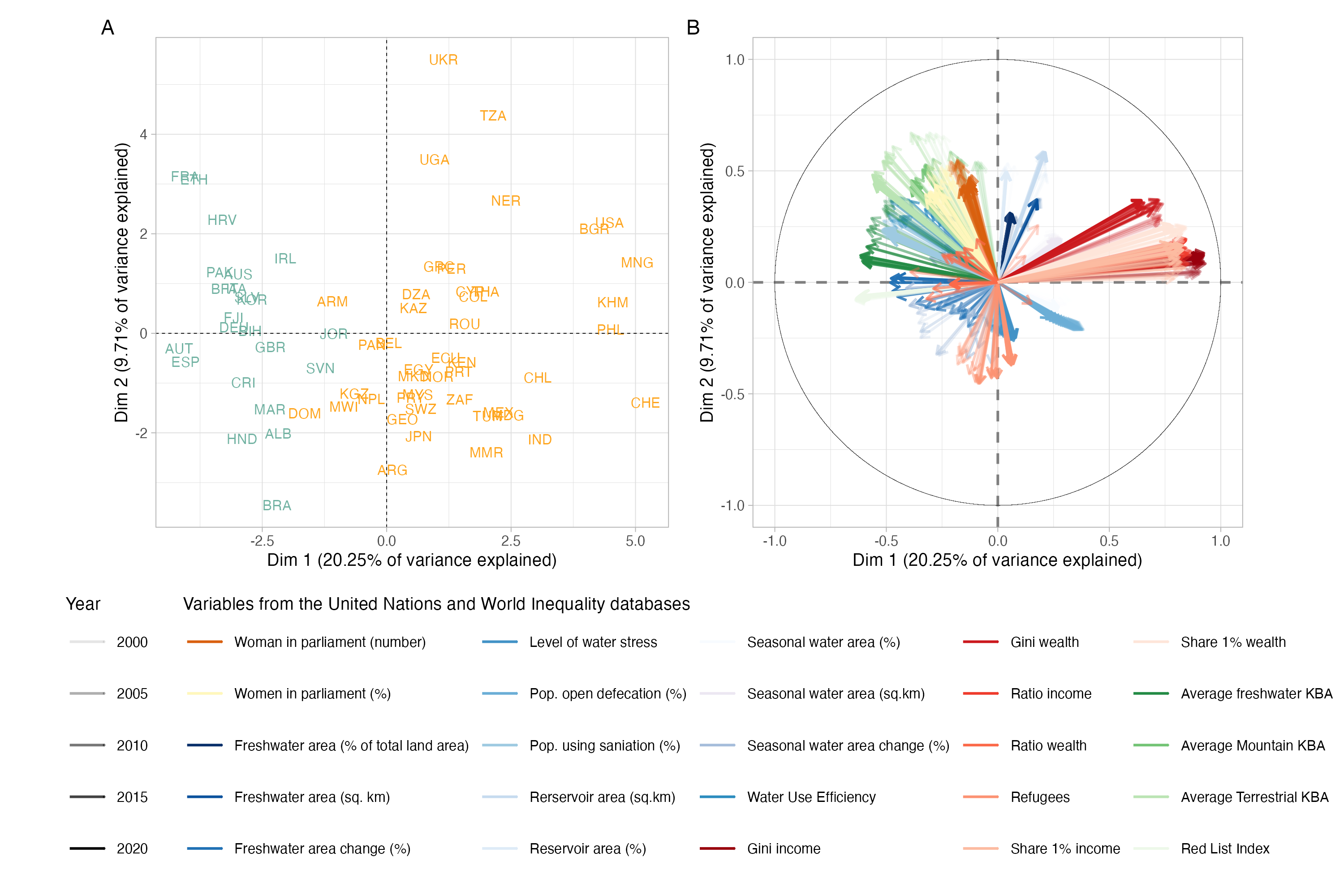}
\caption{\textbf{Multiple factor analysis with United Nations data} The first 10 principal components used in the ordination explain 72\% of the variation, the first two (A) explain 29.9\%. The first 10 components were used to cluster 67 countries resulting in two clusters (A). The correlation circle across explanatory variables is presented in (B) along their loadings on the first two components of the ordination. Variables in the legend are ordered and coloured according to the SDGs used (e.g. reds are inequality, greens are life in land).}
\label{fig:un}
\end{figure*}

Despite the differences in coverage with respect to countries, time, and
indicators tracked, the ordination in both data sets results in two
clusters of countries (Fig \ref{fig:wb}, Fig \ref{fig:un}). For the WB
data set, countries along the first principal component are
differentiated by high levels of economic inequality, high energy
intensity but low carbon emissions (positive values of Dim 1, eg. green
cluster Mexico: MEX or South Africa: ZAF), versus countries with
relatively low gender inequality, high agricultural productivity, high
carbon emissions, high urbanization and internet access (negative values
in Dim 1, amber cluster). Forest related variables have the lowest
loading on the first two components and do not change much over time,
while access to the internet or inclusion of women in leadership roles
have the highest variability over time. The first dimension on the
ordination is best explained by variability in the inequality variables
including women in business, access to internet and electricity, while
the second dimension is best explained by urbanization and carbon
emissions (Fig \ref{fig:SM_WBord}).

The UN dataset offers a similar ordination, where countries with high
values along the first axis and higher values along the second axis have
the highest inequalities (e.g Fig \ref{fig:un}). The inequality
variables are highly correlated but also explain large amounts of the
variance. Contrary to the WB data, here forest related variables do show
variability over time, but variables related to biodiversity loss (Red
list index) or some of the area based indicators for water related SDGs
do not change much over the time period of the data (2000-2021). Places
with lower economic inequality also tend to have better opportunities
for women to participate in political decision making. Interestingly,
high values on the red list index correlate with lower levels of
inequality. The first dimension on the ordination is best explained by
variability in the inequality, while the second dimension is best
explained by key biodiversity areas in terrestrial systems, reservoir
statistics, as well as women participation in parliament (Fig
\ref{fig:SM_UNord}).

\begin{figure*}[ht]
\centering
\includegraphics[width = 6.5in, height = 3.5in]{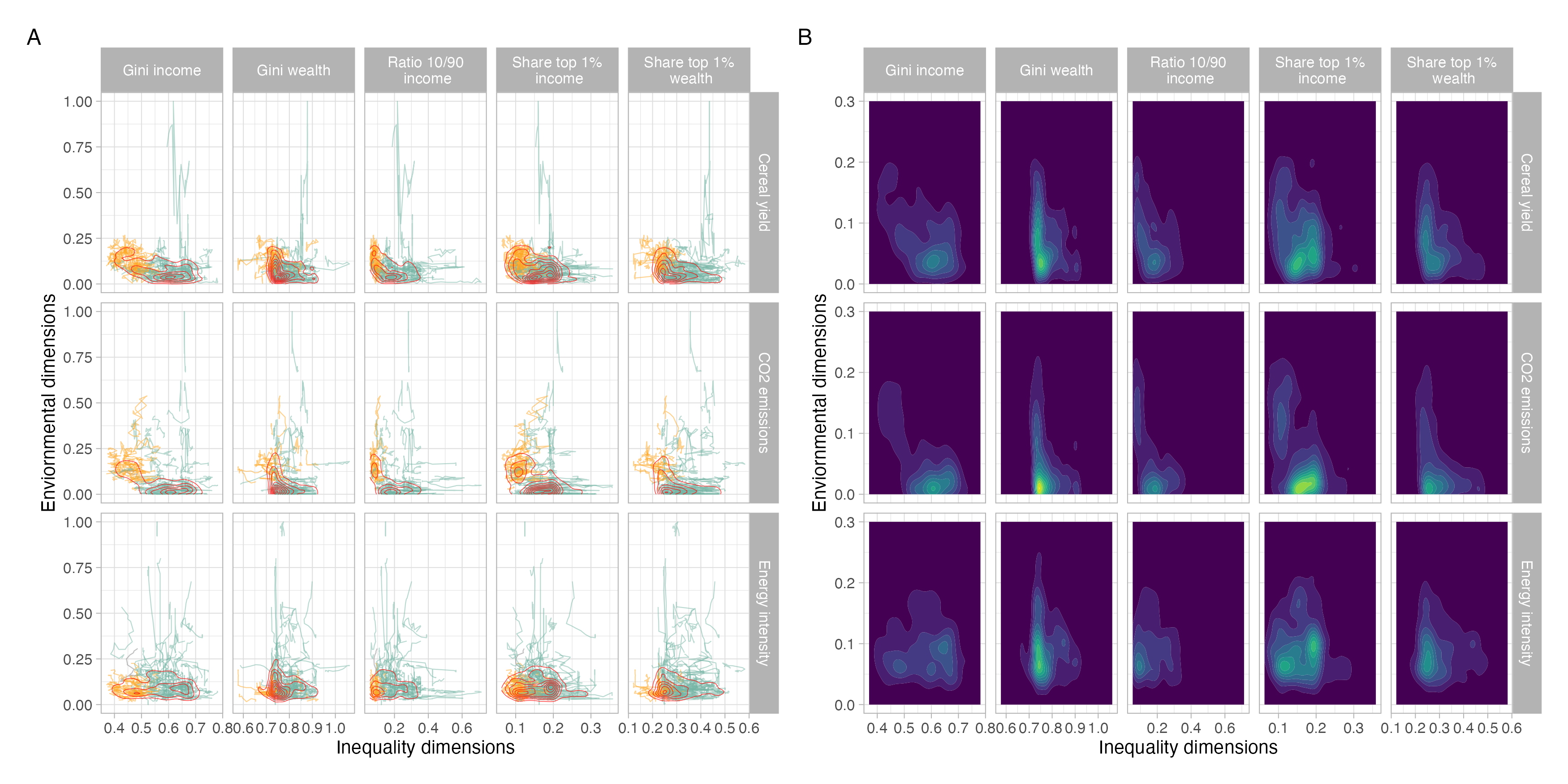}
\caption{\textbf{Inequality regimes} We find bimodal distributions for some dimensions of inequality against environmental factors in the WB dataset. Countries trajectories in A using the same cluster groups as in Fig \ref{fig:un}. Bimodal distributions are more common for the share of the top 1\% than other inequality variables (B). The y-axis has been rescaled to the range 0-1 to ease comparison. Supplementary figure \ref{fig:SM_UNbim} shows a similar plot for the UN data, however lack of variability in the environmental data prevents the identification of multiple modes in the distribution.}
\label{fig:bim}
\end{figure*}

We find evidence of multi-modal distributions in inequality and
environmental variables. A Hartigan's Dip test for unimodality results
on significant p-values for all variables except the Gini in wealth
(ratio of income N = 9145 p \textless{} 2.2e-16, ratio of wealth N =
5766 p = 0.020, share of 1\% wealth N = 11871 p \textless{} 2.2e-16,
Gini of wealth N = 5852 p = 0.113, share of 1\% income N = 19322 p
\textless{} 2.2e-16, Gini of income N = 9258 p = 9.023e-05). Significant
p-values suggest that the distribution is not unimodal, at least
bimodal. This finding supports the hypothesis of the existence of
inequality regimes both in income and wealth, particularly when
inequality is measured as the share of the top 1\% (Fig \ref{fig:bim}).
However, the lack of variability in environmental variables and smaller
sample size in the UN data set prevent us from statistically deriving
modes in the distribution or all our variables (Table \ref{tab:tbl1}).
As a result, we only report bimodal distributions for cereal yields,
carbon emissions, and energy intensity (WB data, Fig \ref{fig:bim}), and
the red list index (UN data, Fig \ref{fig:SM_UNbim}). We also confirm
that, as expected, these inequality regimes are related with the country
typologies identified.

\begin{figure*}[ht] 
\centering 
\includegraphics[width = 6.5in, height = 2in]{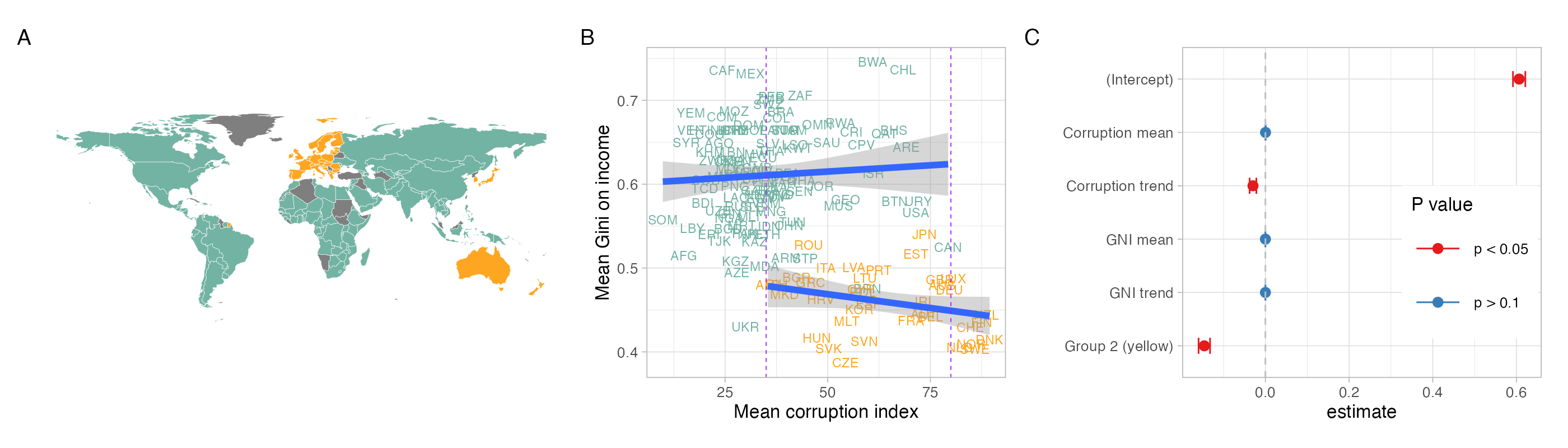} 
\caption{\textbf{Inequality, corruption, and potential hysteresis} We find two inequality regimes that correspond to the country groups previously reported (A). Corruption has been proposed as a generative mechanism of inequality. Here we find support for the idea of corruption being a driver (B) that can tip countries between inequality regimes. For values below 35 or above 80 in the corruption index only one regime exists, in between these values two regimes of inequality exist. A linear regression analysis (N = 148 countries) shows that negative trends on corruption index (lower values means more corruption) increases inequality measured as the mean Gini coefficient on income (C). A regression table for (C) is available on SI Table \ref{tab:reg}.} 
\label{fig:hysteresis} 
\end{figure*}

Using data from the quality of government dataset\textsuperscript{8}, we
find support for the hypothesis that corruption increases
inequality\textsuperscript{24,25}. A linear regression, using the mean
Gini coefficient of income (per country) as dependent variable, shows
the existence of different slopes for the country groups found in the
clustering analysis, whereby the higher the corruption trend (negative
and significant coefficient), the higher the inequality measured as the
mean Gini on income (Fig \ref{fig:hysteresis}). Our results also support
the idea of inequality regimes\textsuperscript{19} in the sense of
finding support for hysteresis between the clusters of countries. There
is a region on the corruption space where the two inequality regimes
co-exist (approximately between 35 and 80). There are not many
transitions between one regime and the other in the historical record,
thus we cannot empirically confirm the existence of a tipping point in
corruption that potentially triggers a country to shift from one
inequality regime to another. Nonetheless, our results provide
provisional evidence for the hypothesis of hysteresis and the existence
of tipping points in inequality\textsuperscript{19}.

\section{Discussion}\label{discussion}

Although nearly all countries have committed to achieving progress on
the SDGs by 2030, it is not clear whether, at the global level, these
goals will be met. Nonetheless, the agenda itself has stimulated
political willingness and policy action in most
countries\textsuperscript{26}. Making progress on SDGs requires
understanding the inevitable synergies and trade-offs between them.
Furthermore, it is clear that socio-economic inequalities can hinder
collective action and other types of political activity conducive for
environmental stewardship\textsuperscript{14,27,28}. Here we address
this challenge directly by investigating SDGs related to inequality and
biosphere stewardship (SDGs 5, 6, 10, 13, 14, and 15) through ordination
methods.

We find evidence of synergies and trade-offs. For example, almost all
inequality metrics (Table \ref{tab:tbl1}) are positively correlated,
except the ratio of wealth which varies less and sometimes negatively
correlated with inequality metrics on income. For example, Sweden
historically has had low income inequality but high wealth inequality,
although the former has been increasing as well. Similarly, gender
equity measured as the share of women in parliament is generally better
in places with lower income inequality but not necessarily low wealth
inequality. Interestingly, countries with high gender equality perform
better in area-based indicators of ecological and biological
conservation. Higher levels of urbanization are correlated with
improvements in access to electricity and internet usage, but also with
higher carbon emissions and lower energy intensity.

Lack of variability in many SDG indicators, in particular related to the
biosphere, questions their utility to track progress towards the SDG
agenda. Biosphere variables are either limited to conservation, or
resource production, consumption, or emissions. Most of the proxies of
ecosystem variables are area-based (e.g.~\% forest area), meaning they
change very little over time. The lack of variability in the SDGs
datasets and a large proportion of missing values (Figs
\ref{fig:SM_WB_env}, \ref{fig:SM_WB_env2}, \ref{fig:SM_UNdata})
compelled us to abandon many indicators. It also questions whether the
indicators currently used are reliable proxies for progress - or the
lack thereof - towards achieving the SDG agenda. If a variable does not
change over time at a scale at which information can feedback to
political decisions, then it inay not be a very useful proxy of progress
towards a desired goal. An important distinction to be made is variables
which do not change but could (e.g.~protected areas), and metrics which
cannot really change fast on the time scale of policy making (e.g.~area
of forest, slow growth rates). Other biodiversity related values such as
cultural values (intrinsic, recreational, spiritual) are not currently
captured by SDG indicators. An interesting avenue for future research is
to design comparable observables that can be monitored by third parties
(not reported by governments) that are sensitive enough to capture
progress or lack of it towards the goals. Some examples include the
essential biodiversity variables initiative advanced by
GeoBON\textsuperscript{29}, the human footprint
index\textsuperscript{30}, recent developments on functional
integrity\textsuperscript{31}, or subnational historical estimates of
inequality\textsuperscript{20,32}. All of them offer time series at
levels of spatiotemporal resolution that enable sub-national monitoring.

Leveraging these subnational data, future research could explore whether
the patterns reported here hold at finer scales. Our national level
analysis of the interactions between inequality and the biosphere falls
short in capturing heterogeneity within countries. Most mechanistic
accounts of the origin of inequality happen at the scale of individuals,
households, or businesses. Preliminary analysis of subnational
inequality shows that different trends and mechanisms could be in place,
calling for different policy interventions depending on
context\textsuperscript{20}. Hence testing for mechanisms would benefit
from a higher resolution in space and time, which we were not able to
test with SDG data.

Despite these limitations, we find evidence for bistability in
inequality, further supporting the hypothesis of inequality
traps\textsuperscript{19}. We also find preliminary evidence for
hysteresis. However the lack of transitions between regimes prevents us
from empirically estimating potential tipping points. To further test
the existence of hysteresis we need a better understanding of the
potential mechanisms at the country scale that might be generating the
alternative regimes, including a controlling parameter at which the
shift from one regime to the other should result in a different break
point depending on the direction of the shift. Here we only investigated
corruption, but other underlying mechanisms
exist\textsuperscript{14,17,18}. Further work can investigate this
hypothesis by analyzing other datasets or longer time series, or by
exploring the plausibility of alternative mechanisms through
modelling\textsuperscript{18}. Biologically-inspired models of
intergenerational wealth accumulation have provided a mechanistic
explanation for the emergence of wealth inequality\textsuperscript{33},
but less is known on which mechanisms might explain income or gender
inequalities, although some hypotheses have been put
forward\textsuperscript{17,18}.

\section{Conclusion}\label{conclusion}

Rising economic inequality is a defining challenge of \(21^{st}\)
century\textsuperscript{34}. We explored potential synergies and
trade-offs between inequality- and biosphere-related Sustainable
Development Goals. We confirm some of the synergies and trade-offs
previously reported, but also show that data gaps, low data quality and
low variability prevent countries from measuring progress towards the
SDG agenda in a meaningful way. We discussed alternative datasets that
could improve independent monitoring, and suggested further studies
investigating whether the patterns reported here hold at subnational
scales. We found support for the hypothesis of inequality traps, regimes
and hysteresis. Preliminarily, we showed that corruption, measured as
corruption perception index, can drive countries between regimes of
inequality. Further testing this theory requires higher resolution data
closer to the scale at which mechanisms are hypothesized.

\subsection{Acknowledgements}\label{acknowledgements}

This work was supported by Formas grant 2020-00454. Our work benefited
from feedback from Kevin Berry, Tracie Curry, Robert Heilmayr, Patrik
Henriksson, Yolanda Lopez-Maldonado, and Per Molander. The findings and
conclusions in this publication are those of the authors and should not
be construed to represent any official USDA or U.S. Government
determination or policy.

\section*{References}\label{references}
\addcontentsline{toc}{section}{References}

\phantomsection\label{refs}

\pagebreak

\section{Supplementary Material}\label{sec:SM}

\renewcommand\thefigure{S\arabic{figure}}
\renewcommand\thetable{S\arabic{table}}
\setcounter{table}{0}
\setcounter{figure}{0}

List of countries analyzed in the UN dataset after removing time series
with too many missing values or too few countries: Albania, Algeria,
Argentina, Armenia, Australia, Austria, Belgium, Bosnia and Herzegovina,
Brazil, Bulgaria, Burkina Faso, Cambodia, Chile, Colombia, Costa Rica,
Croatia, Cyprus, Dominican Republic, Ecuador, Egypt, El Salvador,
Eswatini, Ethiopia, Fiji, Finland, France, Georgia, Germany, Greece,
Honduras, India, Ireland, Italy, Japan, Jordan, Kazakhstan, Kenya,
Kyrgyzstan, Madagascar, Malawi, Malaysia, Mexico, Mongolia, Morocco,
Myanmar, Nepal, Niger, North Macedonia, Norway, Pakistan, Panama,
Paraguay, Peru, Philippines, Portugal, Republic of Korea, Romania,
Slovenia, South Africa, Spain, Switzerland, Thailand, Türkiye, Uganda,
Ukraine, United Kingdom of Great Britain and Northern Ireland, United
Republic of Tanzania and United States of America

List of countries analyzed in the WB dataset after removing time series
with too many missing values or too few countries: Afghanistan, Albania,
Angola, Antigua and Barbuda, Argentina, Armenia, Australia, Austria,
Azerbaijan, Bahamas, The, Bangladesh, Barbados, Belarus, Belgium,
Belize, Benin, Bhutan, Bolivia, Bosnia and Herzegovina, Botswana,
Brazil, Brunei Darussalam, Bulgaria, Burkina Faso, Burundi, Cabo Verde,
Cambodia, Cameroon, Canada, Central African Republic, Chad, Chile,
China, Colombia, Comoros, Congo, Dem. Rep., Congo, Rep., Costa Rica,
Cote d'Ivoire, Croatia, Cyprus, Czech Republic, Denmark, Djibouti,
Dominica, Dominican Republic, Ecuador, Egypt, Arab Rep., El Salvador,
Eritrea, Estonia, Eswatini, Ethiopia, Fiji, Finland, France, Gabon,
Gambia, The, Georgia, Germany, Ghana, Greece, Grenada, Guatemala,
Guinea, Haiti, Honduras, Hungary, India, Indonesia, Iran, Islamic Rep.,
Ireland, Israel, Italy, Jamaica, Japan, Jordan, Kazakhstan, Kenya,
Korea, Rep., Kuwait, Kyrgyz Republic, Lao PDR, Latvia, Lebanon, Lesotho,
Libya, Lithuania, Luxembourg, Madagascar, Malawi, Maldives, Mali, Malta,
Mauritania, Mauritius, Mexico, Micronesia, Fed. Sts., Moldova, Mongolia,
Morocco, Mozambique, Myanmar, Namibia, Nepal, Netherlands, New Zealand,
Nicaragua, Niger, Nigeria, North Macedonia, Norway, Oman, Pakistan,
Panama, Papua New Guinea, Paraguay, Peru, Philippines, Poland, Portugal,
Qatar, Romania, Russian Federation, Rwanda, Sao Tome and Principe, Saudi
Arabia, Senegal, Sierra Leone, Slovak Republic, Slovenia, Solomon
Islands, Somalia, South Africa, Spain, Sri Lanka, St.~Vincent and the
Grenadines, Suriname, Sweden, Switzerland, Syrian Arab Republic,
Tajikistan, Tanzania, Thailand, Togo, Trinidad and Tobago, Tunisia,
Uganda, Ukraine, United Arab Emirates, United Kingdom, United States,
Uruguay, Uzbekistan, Vanuatu, Venezuela, RB, Vietnam, Yemen, Rep.,
Zambia and Zimbabwe

\begin{figure*}[ht]
\centering
\includegraphics[width = 6in, height = 4in]{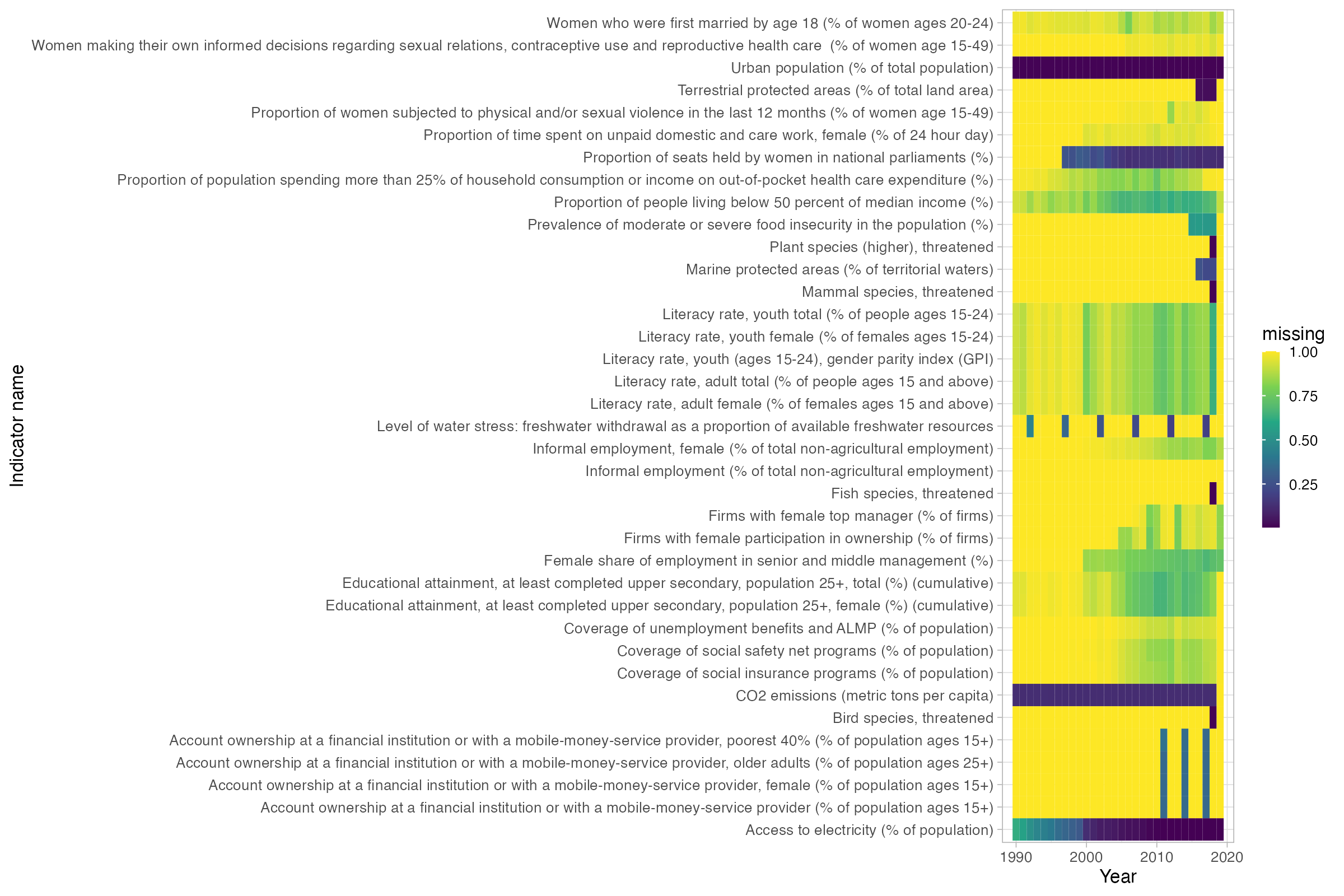}
\caption{\textbf{World Bank environmental SDGs} }
\label{fig:SM_WB_env}
\end{figure*}
\begin{figure*}[ht]
\centering
\includegraphics[width = 6in, height = 6in]{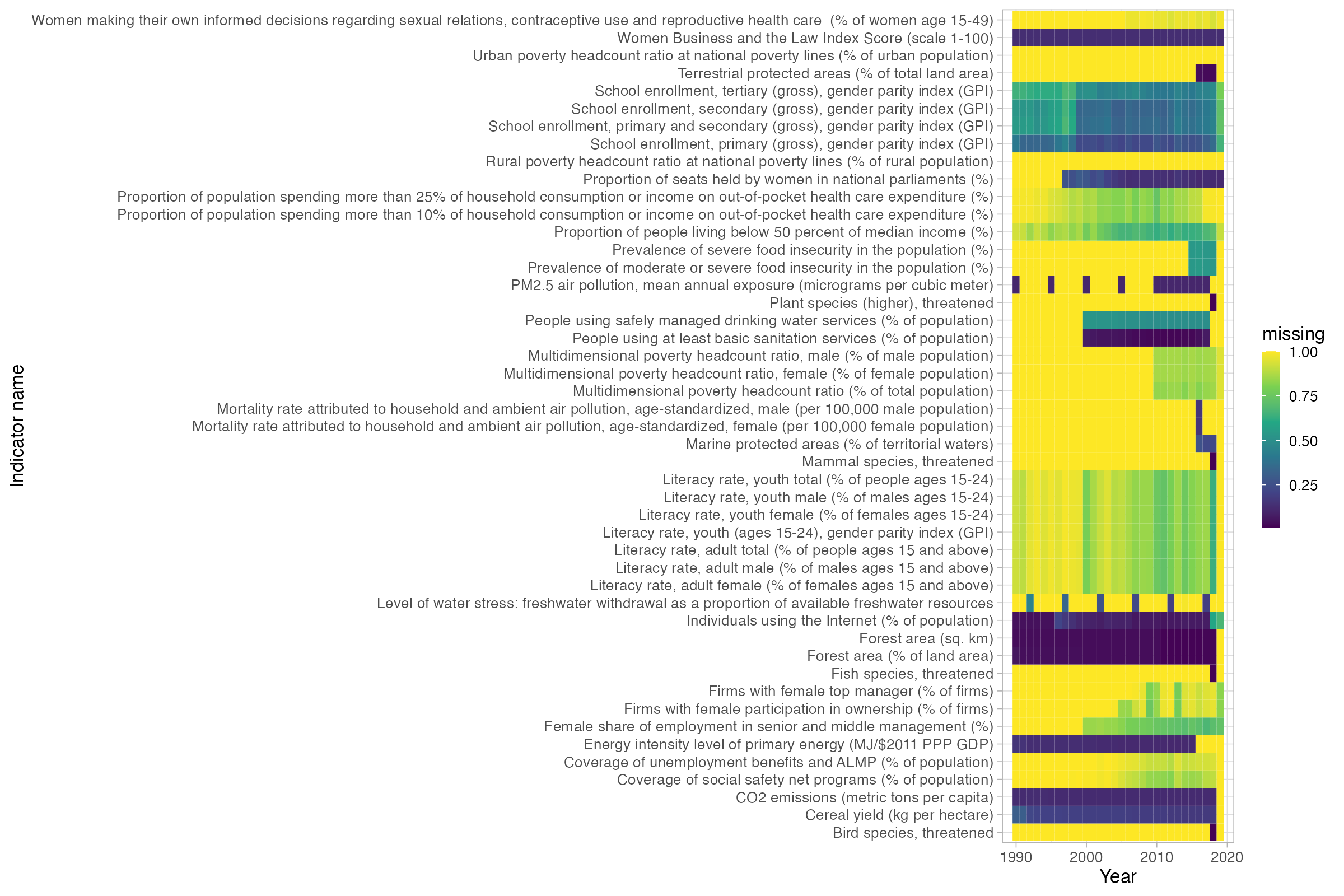}
\caption{\textbf{World Bank environmental SDGs additional variables} }
\label{fig:SM_WB_env2}
\end{figure*}
\begin{figure*}[ht]
\centering
\includegraphics[width = 6in, height = 6in]{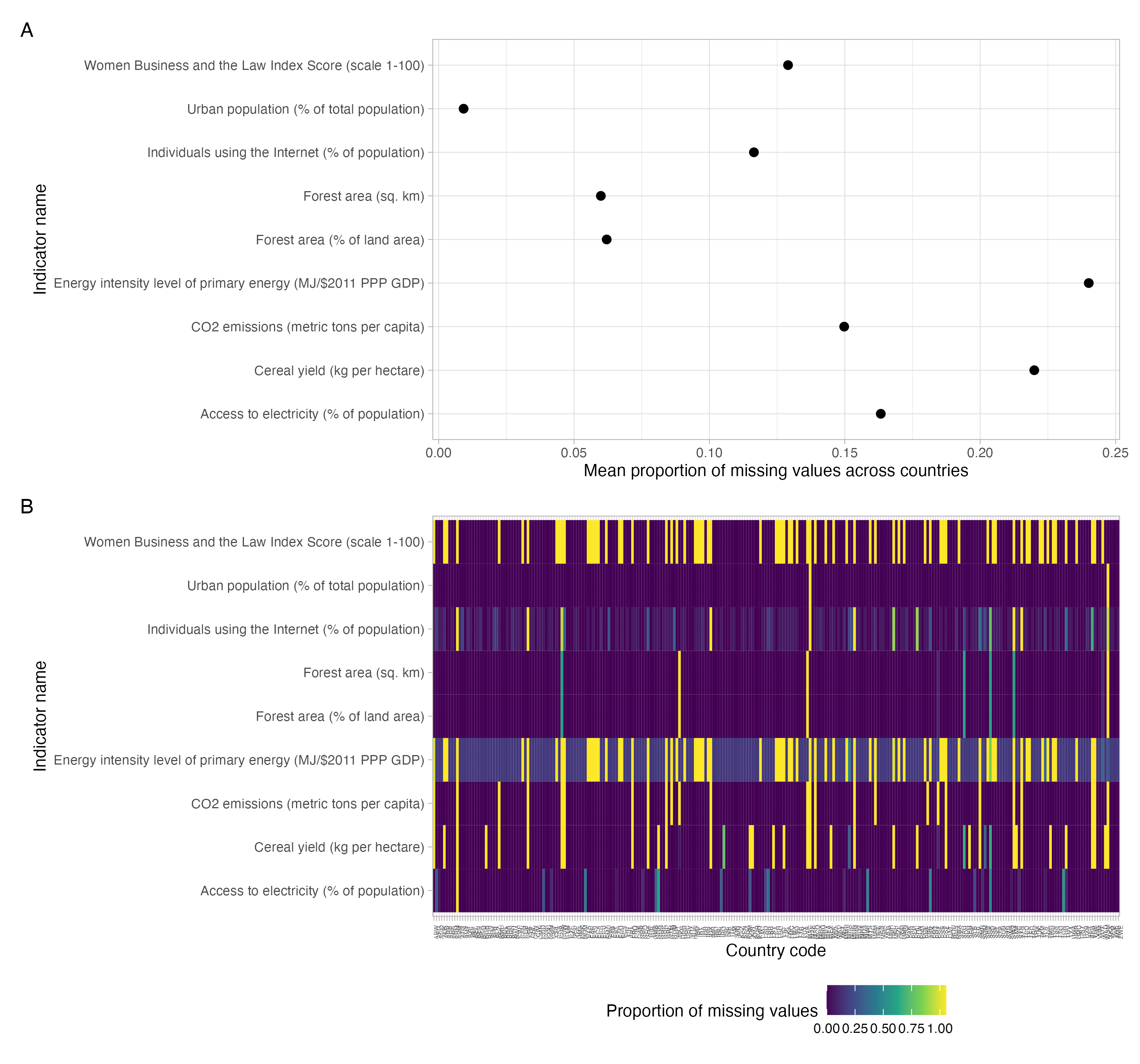}
\caption{\textbf{World Bank selected variables} }
\label{fig:SM_WB_select}
\end{figure*}
\begin{figure*}[ht]
\centering
\includegraphics[width = 6in, height = 5in]{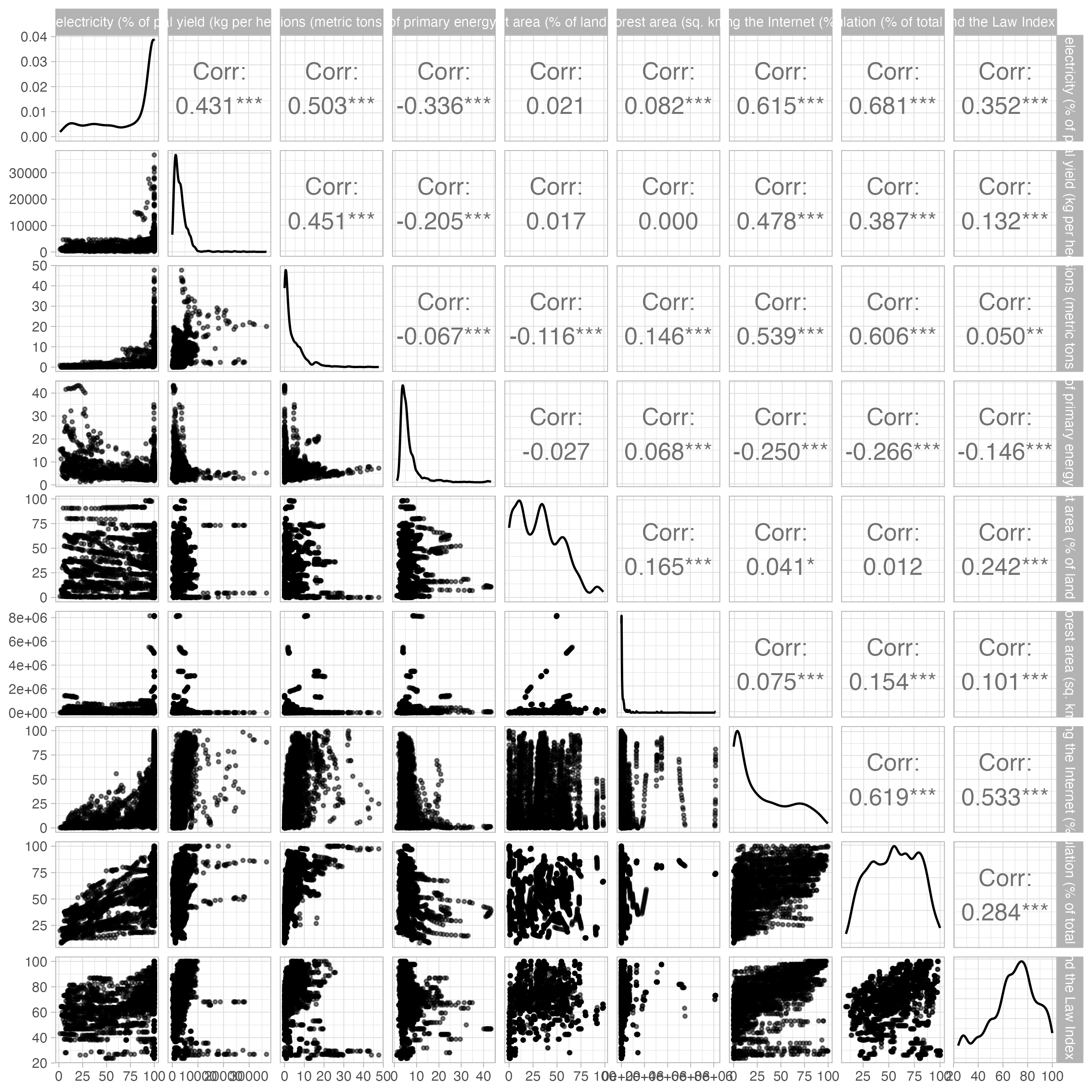}
\caption{\textbf{Correlogram of World Bank selected variables} }
\label{fig:SM_WBcorr}
\end{figure*}
\begin{figure*}[ht]
\centering
\includegraphics[width = 6in, height = 5in]{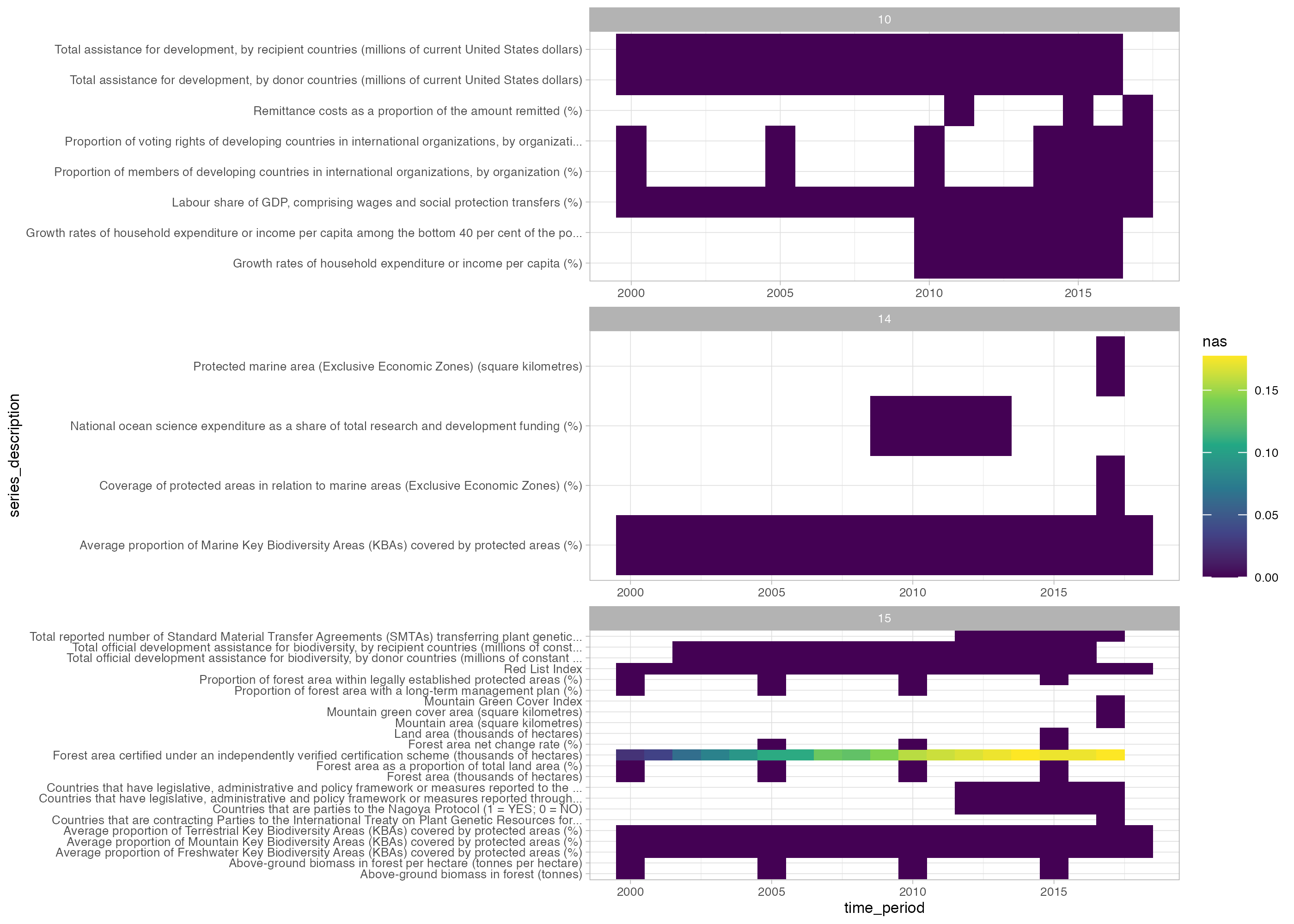}
\caption{\textbf{United Nations SDGs 10, 14 and 15 dataset} }
\label{fig:SM_UNdata}
\end{figure*}
\begin{figure*}[ht]
\centering
\includegraphics[width = 6in, height = 5in]{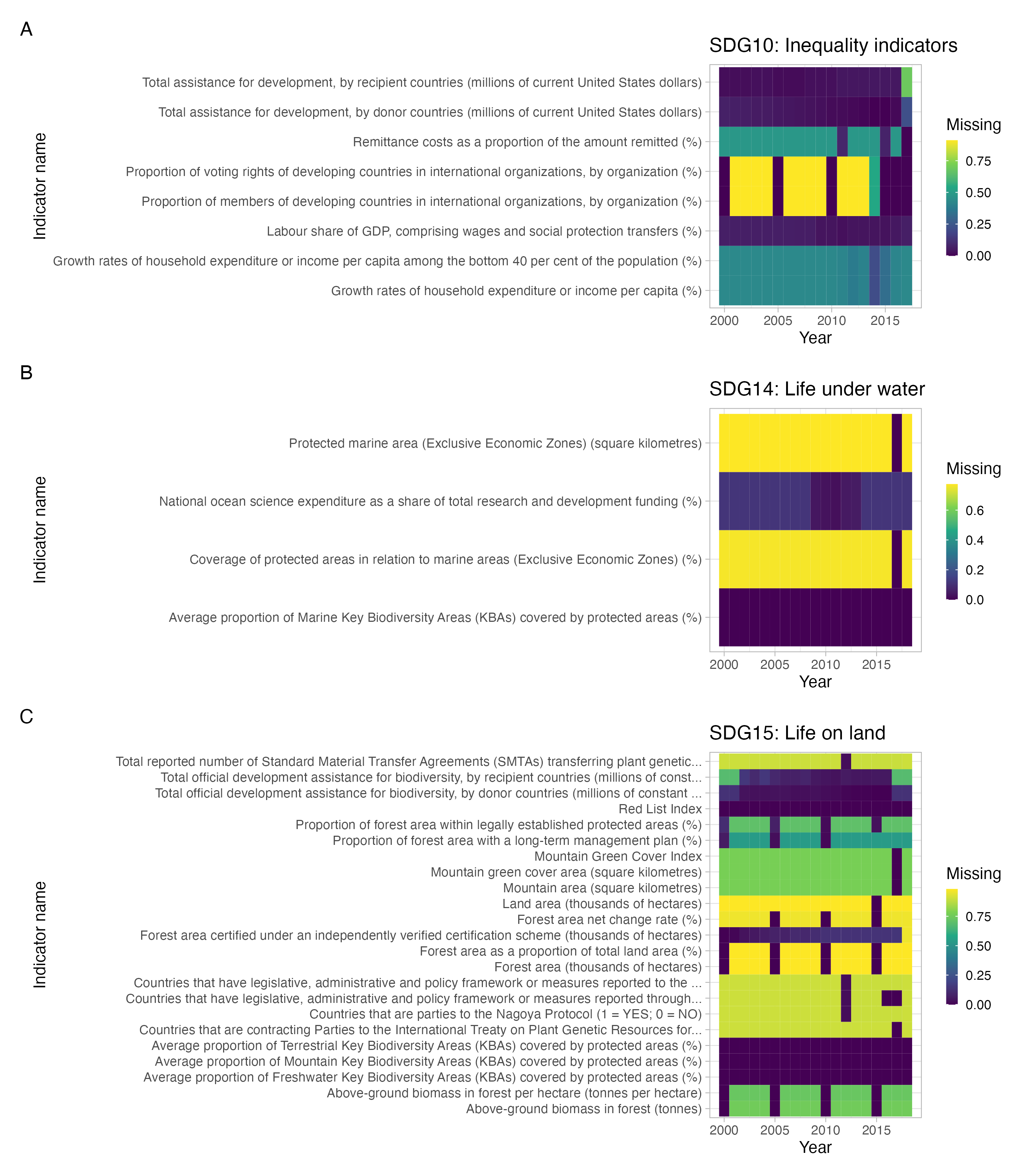}
\caption{\textbf{UN selected variables} }
\label{fig:SM_UNselect}
\end{figure*}
\begin{figure*}[ht]
\centering
\includegraphics[width = 6in, height = 5in]{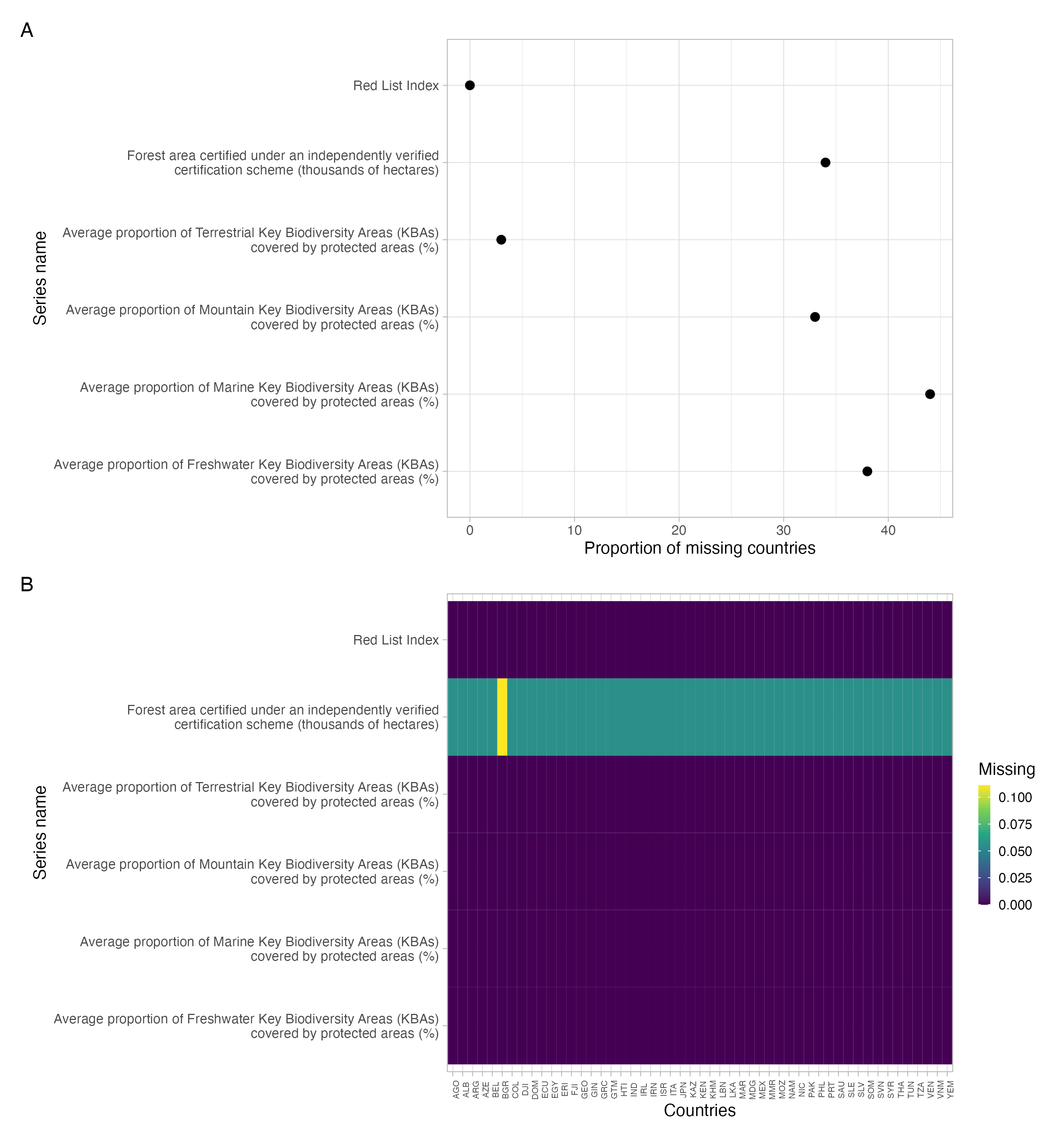}
\caption{\textbf{UN reduced variables} }
\label{fig:SM_UNred}
\end{figure*}
\begin{figure*}[ht]
\centering
\includegraphics[width = 6in, height = 5in]{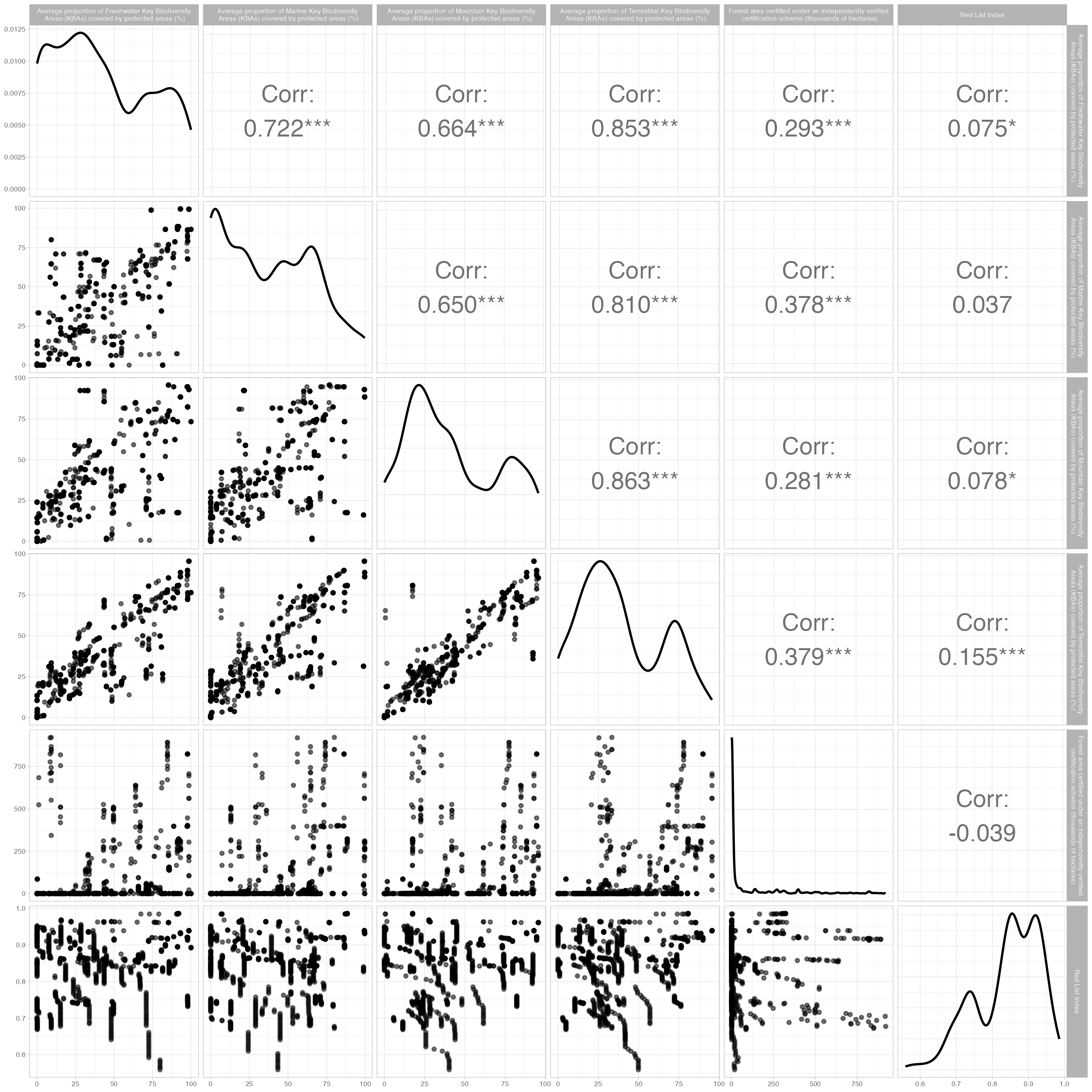}
\caption{\textbf{Correlogram of United Nations selected variables} }
\label{fig:SM_UNcorr}
\end{figure*}
\begin{figure*}[ht]
\centering
\includegraphics[width = 6in, height = 5in]{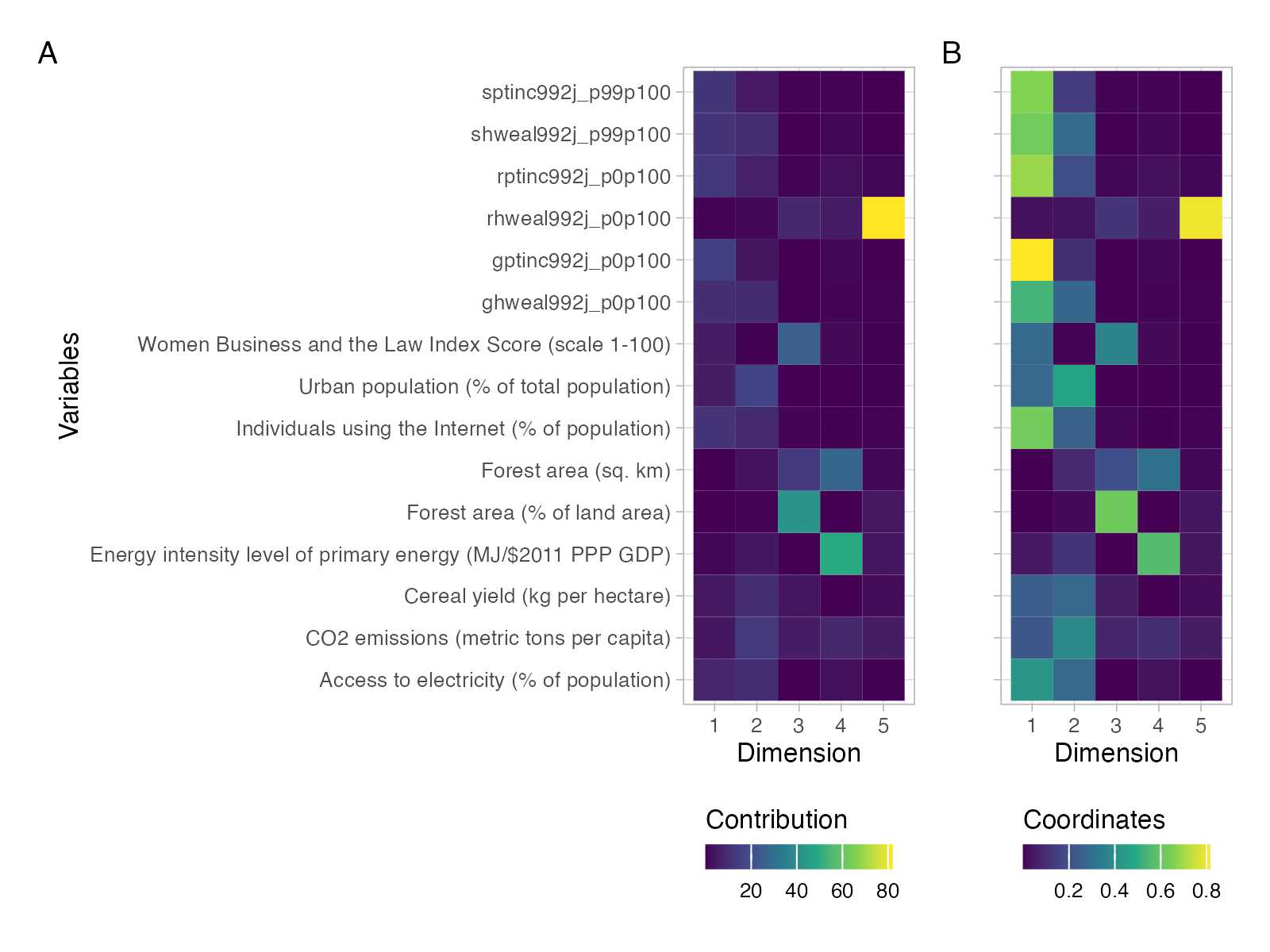}
\caption{\textbf{Variable importance in ordination on WB data} }
\label{fig:SM_WBord}
\end{figure*}
\begin{figure*}[ht]
\centering
\includegraphics[width = 6in, height = 5in]{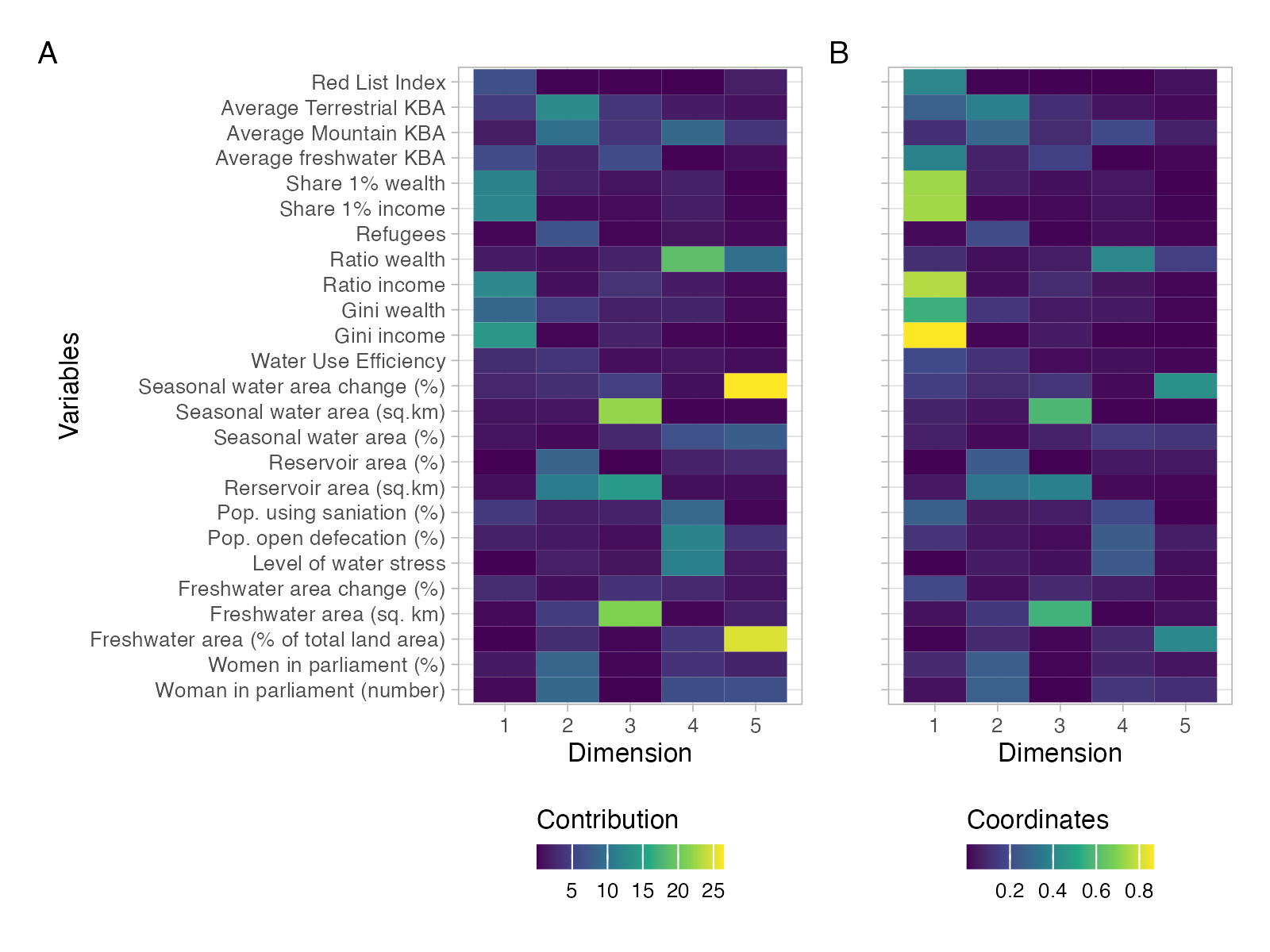}
\caption{\textbf{Variable importance in ordination on UN data} }
\label{fig:SM_UNord}
\end{figure*}
\begin{figure*}[ht]
\centering
\includegraphics[width = 4in, height = 3.5in]{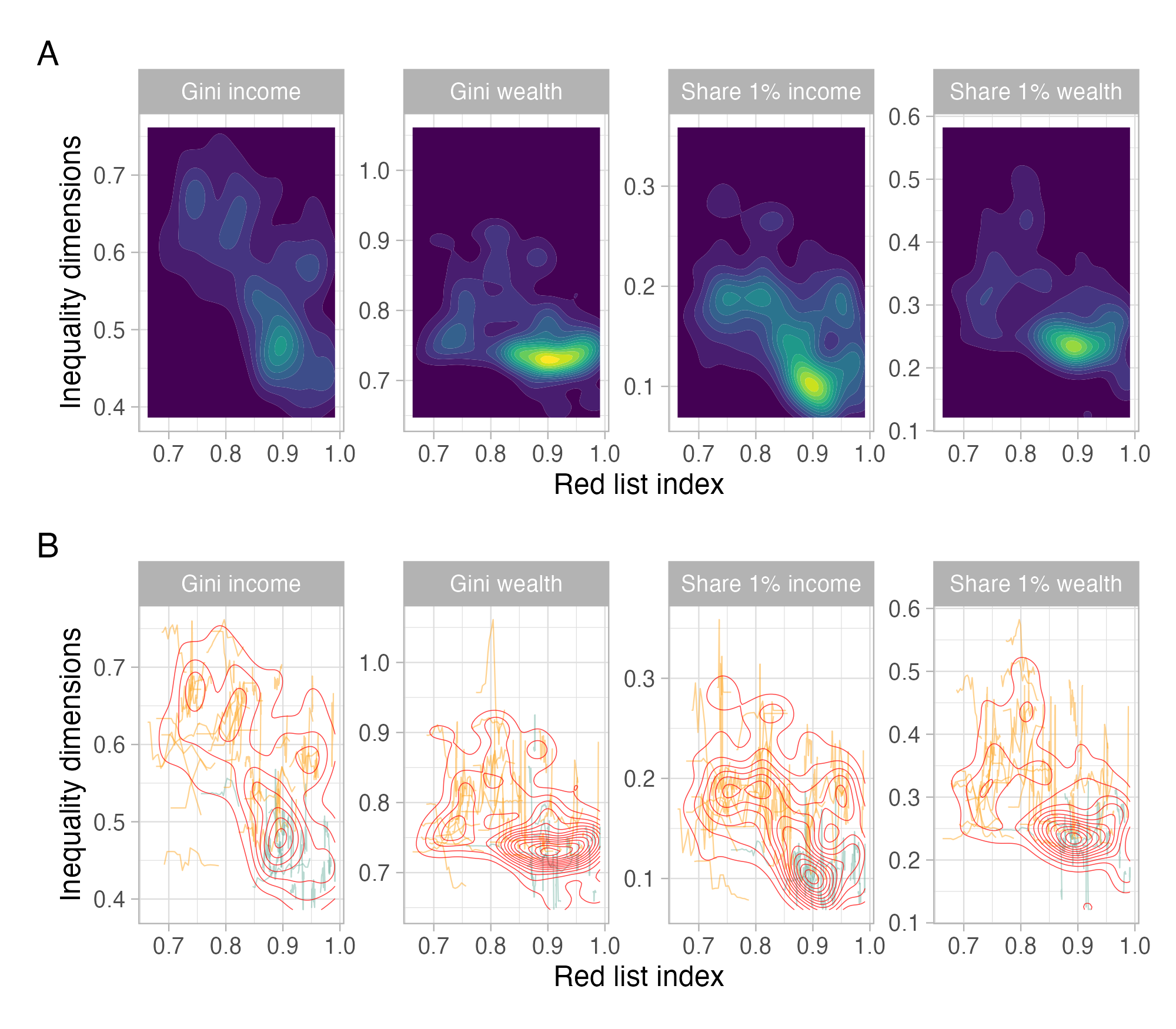}
\caption{\textbf{Bimodality in UN data} The quality of the data does not allow for detection of bimodality in most UN variables. The lack of variation in biosphere related data does not allow the identification of areas in the parameter space where trajectories converge, except for the red list index, where we find multimodal distributions }
\label{fig:SM_UNbim}
\end{figure*}

\begin{table}[!htbp] \centering 
  \caption{Regression table for Figure 4C. Mean inequality is regressed against mean and trend of corruption index for 148 countries after controlling for gross national income (mean and trend)} 
  \label{tab:reg} 
\begin{tabular}{@{\extracolsep{5pt}}lc} 
\\[-1.8ex]\hline 
\hline \\[-1.8ex] 
 & \multicolumn{1}{c}{\textit{Dependent variable:}} \\ 
\cline{2-2} 
\\[-1.8ex] & Mean Gini on income \\ 
\hline \\[-1.8ex] 
 Mean corruption index & 0.0003 \\ 
  & (0.0004) \\ 
  & \\ 
 Trend on corruption index & $-$0.030$^{***}$ \\ 
  & (0.008) \\ 
  & \\ 
 Mean GNI & $-$0.00000 \\ 
  & (0.00000) \\ 
  & \\ 
 Trend GNI & 0.00000 \\ 
  & (0.00001) \\ 
  & \\ 
 Group 2: yellow cluster & $-$0.146$^{***}$ \\ 
  & (0.014) \\ 
  & \\ 
 Constant & 0.606$^{***}$ \\ 
  & (0.015) \\ 
  & \\ 
\hline \\[-1.8ex] 
Observations & 148 \\ 
R$^{2}$ & 0.624 \\ 
Adjusted R$^{2}$ & 0.610 \\ 
Residual Std. Error & 0.054 (df = 142) \\ 
F Statistic & 47.071$^{***}$ (df = 5; 142) \\ 
\hline 
\hline \\[-1.8ex] 
\textit{Note:}  & \multicolumn{1}{r}{$^{*}$p$<$0.1; $^{**}$p$<$0.05; $^{***}$p$<$0.01} \\ 
\end{tabular} 
\end{table}

\end{document}